\documentclass[12pt]{article}
\usepackage{graphicx}
\begin{document}
\begin{center}
{\Large\bf  Constraining a scalar field dark energy with variable equation of state for matter }\\[20mm]
A. Sil \footnote{St.Paul's C. M. College, 33/1 Raja Rammohan Sarani, Kolkata 700 009, India;\\
\indent email: amitavadrsil@rediffmail.com}and S. Som \footnote{Meghnad Saha Institute of Technology, Nazirabad, East Kolkata Township, Kolkata 700 107, India;\\ \indent email: sumitsom79@yahoo.com}\\
{\em Relativity and Cosmology Research Centre,\\Department of Physics, Jadavpur University,\\
Kolkata - 700 032, India.} \\[20mm]
\end{center}

\pagestyle{myheadings}
\newcommand{\be}{\begin{equation}}
\newcommand{\ee}{\end{equation}}
\newcommand{\bea}{\begin{eqnarray}}
\newcommand{\eea}{\end{eqnarray}}

\begin{abstract}
The red-shift $z_{eq}$, marking the end of radiation era and the beginning of matter-dominated era, can play an important role to reconstruct dark-energy models. A variable equation of state for matter that can bring a smooth transition from radiation to matter-dominated era in a single model is proposed to estimate $z_{eq}$ in dark energy models and hence its viability. Two one-parameter models with minimally coupled scalar fields playing the role of dark energy are chosen to demonstrate this point. It is found that for desired late time behavior of the models, the estimated value of $z_{eq}$ is highly sensitive on the value of the parameter in each of these models.
\end{abstract}

\vspace{0.5cm}
Key Words: Cosmology; Scalar field, Dark energy.
\vspace{0.5cm}

\section{Introduction}
It has been confirmed by independent observation data based on different sources such as type Ia Supernovae (SN Ia) \cite{riess}, Cosmic Microwave Background radiation(CMBR)\cite{cmbr} and Baryon Acoustic Oscillation \cite{bao}that about 70\% of the energy density of the present universe consists of `Dark Energy'. Unlike ordinary matter the dark energy has a repulsive effect and its dominance accounts for the observed late-time cosmic acceleration. All the research efforts in the area of cosmology over last decade was mainly focussed to understand the origin and nature of dark energy.
\par The most natural and simplest choice as a candidate for dark energy could have been the Cosmological Constant $\Lambda$ with equation of state $\omega _{DE}=-1$. But if one believes that vacuum energy is the origin of it then one fails to find any mechanism to obtain a value of $\Lambda$ that is 120 orders of magnitude less than the theoretical prediction to be consistent with observation. Attempts were made by introducing the concept of varying $\Lambda$ to explain observation but with same equation of state $\omega _{DE}=-1$\cite{varying lambda}. Possibilities of evolving $\omega _{DE}$ were also explored in many dynamical dark energy models. Primary candidates in this category are scalar field models such as Quintessence\cite{quintessence} and K-essence \cite{k essence}. In spite of wide variety of variation in $\omega _{DE}$ and having features as demanded by observation in those models, it is not sufficient to prefer such models over the $\Lambda$ Cold-Dark-Matter ($\Lambda$CDM) model. Also in order to produce late-time acceleration, the potential in those models are so flat that the field mass becomes extremely small \cite{field mass}. However in the framework of particle physics the construction of scalar field dark energy models are most natural and not ruled out. There exists other classes of `modified matter' dynamical dark energy models, such as Chaplygin gas\cite{chaplygin} and also many `modified General Relativity' theories such as $f(R)$ gravity\cite{f R}, Scalar-Tensor Theory\cite{st} inspired models and Braneworld models\cite{brane} in the list. A few comprehensive and recent review articles on dark energy may be of interest in this regard\cite{darkenergy}.
\par Most of these models are usually studied only in the matter dominated phase assuming the presence of some kind of dark energy to explain the transition from deceleration to acceleration. This is due to the fact that radiation era was very short lived and dominated the universe in its early phase of evolution whereas acceleration is a more recent phenomena which has occurred very late in the matter dominated epoch. The viability of such models are usually checked by comparing the variation of physical parameters, e.g.  deceleration parameter $q$ with the red-shift $z$ obtained theoretically with that from observations. A good estimation of $z=z_t$, the redshift marking the transition from deceleration to acceleration, is available from observation. But observational data are limited over a small range of $z$ values around $z_t$. Hence many models that fit observational data are found to behave quite differently outside this range. To choose among these models we need at least another event that occurred outside this range of $z$, preferably in distant past. We point out that a very good estimate of $z=z_{eq}$, the redshift when matter density was equal to radiation density, is at our disposal. Thus instead of a single point we now have two well separated points through which the $q$ vs. $z$ curve for any viable cosmological model should pass. Such a model would be more reliable as its viability could be ensured over a large range of $z$ values. But then to estimate $z_{eq}$ we must have both radiation and matter in a single model. In an earlier work \cite{paper1}we have proposed a variable equation of state that mimics radiation in the past and pressureless dust in later course of evolution. To establish our point we also demonstrated in that work how one can choose a better model among two varying $Lambda$ models.
\par In the next section we introduce the proposed variable equation of state for the cosmological fluid for completeness. In subsequent sections we examine two scalar field models with such equation of state to demonstrate our point.

\section{Radiation and matter in a single model}\label{2}

Let us begin with a Robertson-Walker spacetime with flat spatial geometry following strong observational evidences\cite{zhao}. In Friedmann models this homogeneous and isotropic universe is filled with an ideal fluid. The equation of state for this fluid is either that of radiation ($p_{r}=\frac{1}{3}\rho_{r}$) or that of pressureless dust ($p_{m}=0$).In a standard cosmological model such universe with no dark energy radiation dominates for roughly the first 2000 years of evolution and matter dominates there-after. Thus one needs to join two different cosmological models with different equations of state for matter at some particular time to explain the expansion history of the universe.
\par Attempts were made to describe the entire evolution in a single model by Chernin\cite{chernin}, McIntosh \cite{mcintosh}, Landsberg and Park\cite{lb}, Jacobs\cite{jacobs} and Cohen\cite{cohen} in late sixties of the last century. In those models the homogeneous and isotropic universe is assumed to be filled with a perfect fluid which has two components {\it viz.} radiation and matter. Assuming no interaction between radiation and matter, the radiation energy density $\rho_{r}$ decreases with time in the usual manner $\rho_{r} \sim a^{-4}$ while the matter density falls as $\rho_{m} \sim a^{-3}$. Thus the total energy density $\rho$ and the total pressure $p$ of the fluid may be written as
\bea
\rho = \rho_{r} + \rho_{m} = \rho_{r0}(a_{0}/a)^{4}+\rho_{m0}(a_{0}/a)^{3}\\
p = p_{r}+p_{m}=\frac{1}{3}\rho_{r0}(a_{0}/a)^{4}
\eea
where the subscript `$0$' indicates the value at present epoch. We thus find that $\rho$ and $p$ maintains an equation of state
\be\label{eos}
f=\frac{p}{\rho}=\frac{1}{3}\left[ 1+( \rho_{m0}/\rho_{r0})(a/a_{0})\right]^{-1}=\frac{1}{3}\left[ 1+\frac{a}{a_{eq}}\right]^{-1}  \label{eos1}
\ee
where $a_{eq}$ is defined as the scale when $\rho_{r}$ and $\rho_{m}$ are equal {\it i.e.}
\be
\rho_{r0}(a_{0}/a_{eq})^{4}=\rho_{m0}(a_{0}/a_{eq})^{3}.
\ee
One may see that for small values of `$a$' the equation of state approximates to that of pure radiation while for $a>>a_{eq}$ the pressure becomes negligible and mimics the matter dominated epoch.

\par We propose the cosmological fluid to obey the above equation of state. With such a fluid in presence of dark energy it becomes possible to inspect whether a model is capable of producing the desired or observed values for both $z_{eq}$ and $z_{t}$ simultaneously and hence easier to judge its viability. In an earlier attempt\cite{paper1} we demonstrated this point with varying $\Lambda$ models. In this present paper, we shall consider models with scalar field playing the role of `dark energy' to demonstrate this point.

\section{Field Equations}
For a homogeneous isotropic flat Robertson-Walker metric
\begin{equation}\label{metric}
    ds^2=-dt^2 +a^2(dr^2+r^2d\theta^2+r^2\sin^2\theta d\phi^2)
\end{equation}
the Einstein's field equations for a two component fluid are
\begin{eqnarray}
  3H^2 &=& \rho+\rho_{\phi} \label{feq1}\\
 \mbox{and}\hspace{1cm}2\dot{H}+3H^2 &=& -(p+p_{\phi})\label{feq2}
\end{eqnarray}
where an over dot denotes derivative with respect to time. Here $\rho$ and $p$ are density and pressure of the background cosmological fluid whereas $\rho_{\phi}$ and $p_{\phi}$ respectively denote the dark energy density and pressure. Throughout this article suffix $\phi$ denotes the `dark energy' part of the corresponding quantity. $H$ is the usual Hubble's parameter defined as $H \equiv \dot{a}/a$. The above field equations can also be written as
\begin{eqnarray}
  1 &=& \Omega+\Omega_{\phi} \\
 \mbox{and}\hspace{1cm}\frac{1}{3}(2q-1) &=& \omega\Omega+\omega_{\phi}\Omega_{\phi}
\end{eqnarray}
where deceleration parameter $q$, equation of state $\omega $ and density parameter $\Omega$ are defined in the usual way
\begin{eqnarray}
q &\equiv & -\ddot{a}/aH^2 \nonumber \\
\omega &\equiv & p/\rho \nonumber \\
\mbox{and} \hspace{1cm}\Omega & \equiv & \rho/3H^2. \nonumber
\end{eqnarray}
If we now assume that dark energy is due to a time dependent scalar field minimally coupled with gravity then we could start from an action
\begin{equation}\label{action}
    \mathcal{A}=\int d^4x\sqrt{-g}\left[\frac{R}{16\pi G}+\frac{1}{2}\phi _{,\mu}\phi^{,\mu}-V(\phi)+\mathcal{L}\right]
\end{equation}
to get the same field equations (\ref{feq1}) and (\ref{feq2})with
\begin{eqnarray}
  \rho_{\phi} &=& \frac{1}{2}\dot{\phi}^2+V(\phi) \\
  \mbox{and}\hspace{1cm}p_{\phi} &=& \frac{1}{2}\dot{\phi}^2-V(\phi).
\end{eqnarray}
Since there are altogether five unknown functions $viz.$ $H, \phi , V(\phi), \rho$ and $p$, to solve the set of two field equations we must now make three assumptions.
\begin{enumerate}
\item If we are interested in investigation of late time evolution of the universe we may assume $p=p_{m}=0$ indicating matter in the form of dust. In Model 1 and Model 2 we have only matter in the form of dust. However in Model 1A and Model 2A radiation is also introduced so that $p\neq 0$
\item We may also assume that there exists no interaction between background fluid and dark energy leading to the conservation equation
\begin{equation}\label{conservation}
    \dot{\rho}+3H(\rho+p)=0
\end{equation}
and the wave equation for the scalar field
\begin{equation}\label{wave}
    \ddot{\phi}+3H\dot{\phi}+V'(\phi)=0
\end{equation}
where a prime over head denotes derivative with respect to $\phi$.
\item In the remaining assumption we constrain the scalar field behavior to induce late time acceleration. We make two choices with the first one implemented in Model 1 and Model 1A. The second choice is applied in Model 2 and Model 2A in the subsequent sections.
\end{enumerate}

\section{Model 1}
In this model we assume matter in the form of dust and hence $p=0$. The conservation equation (\ref{conservation}) then immediately leads to
\begin{equation}\label{assump1}
    \rho=A_1 a^{-3}.
\end{equation}
The scalar field is assumed to behave as
\begin{equation}
    \phi=\beta \ln a.
\end{equation}
The assumption is quite similar to the standard choice of scalar field $\phi=\beta \ln t$ to drive power law expansion giving rise to exponential nature in potential.
The assumption (\ref{assump1}) helps to solve the field equations to give
\begin{equation}
  V(\phi) = \frac{A_1\beta^2}{2(3-\beta^2)}e^{-\frac{3}{\beta}\phi}+Ce^{-\beta\phi}.
\end{equation}
Potentials of exponential nature are quite common and well known to produce acceleration in dark energy models\cite{exp pot}
\par
\begin{figure}
    \begin{center}
     \includegraphics[width=6cm]{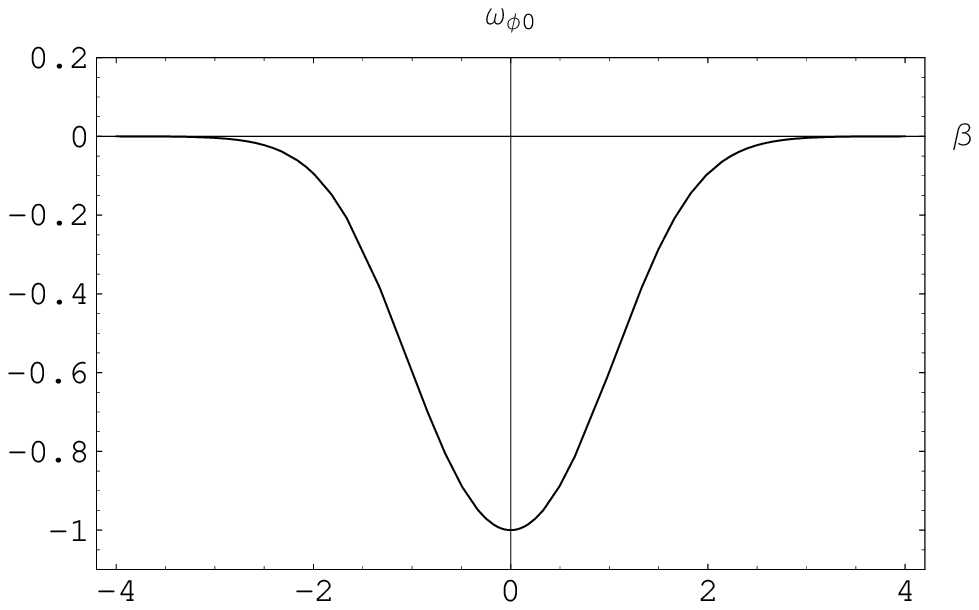}
     \includegraphics[width=6cm]{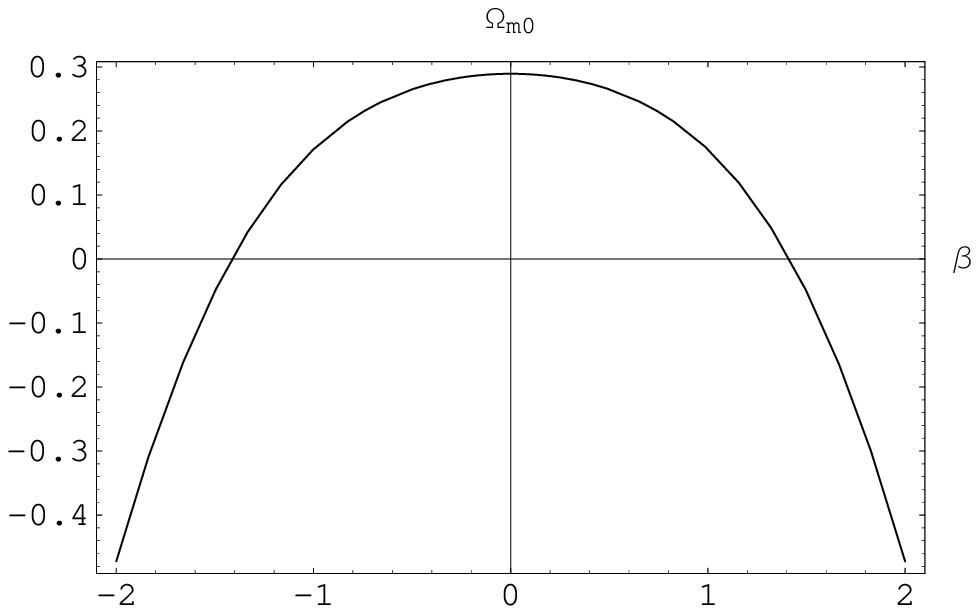}\\
     \includegraphics[width=6cm]{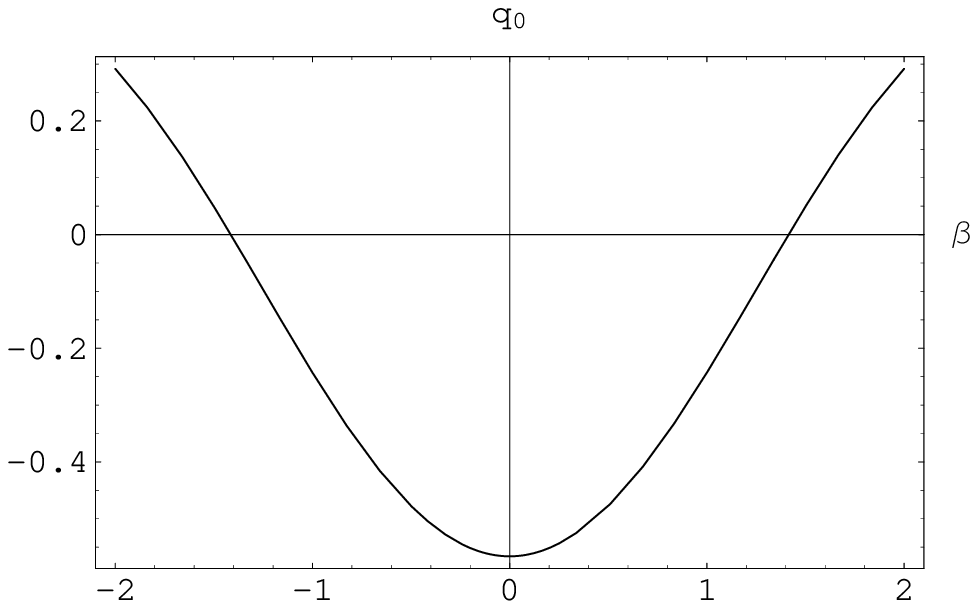}
     \includegraphics[width=6cm]{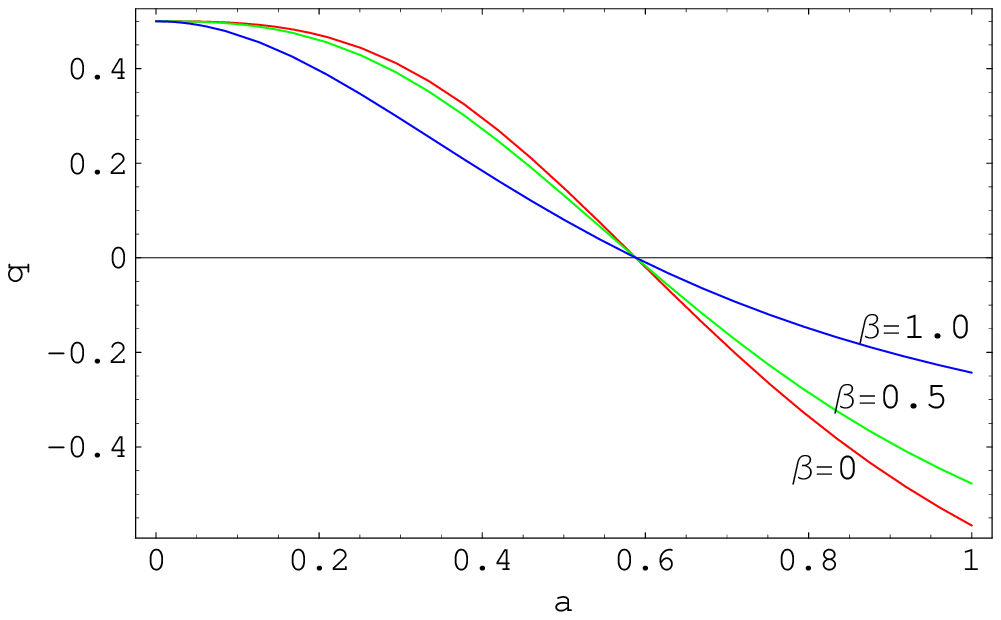}
  \caption{Variation of $\omega_{\phi0}$, $\Omega_{0}$ and $q_0$ with $\beta$ are shown for $a_{0}/a_{t} =1.7$ in Model 1 along with the nature $q$ vs. $a$ curves for different values $\beta $}\label{q_0vsbM1}
  \end{center}
\end{figure}

 If $a_{t}$ defines the scale when universe transits from deceleration to acceleration phase the deceleration parameter $q$ must experience a signature flip at $a=a_{t}$. This helps to determine the constant of integration $C$ to be
\[
    C=\frac{A_1(6-\beta^2)}{2(3-\beta^2)(2-\beta^2)}a_{t}^{\beta^2-3}.
\]
Thus the potential written as a function of $a$ is given by
\begin{equation}
V(a) = \frac{A_1}{2(3-\beta^2)}a^{-3}\left[\beta^2+(6-\beta^2)F\right].
\end{equation}
where $F$ is defined as
\[
F\equiv(2-\beta^2)^{-1}\left(\frac{a}{a_{t}}\right)^{3-\beta^2}
\]
Other physical parameters are listed below as functions of $a$.
\begin{eqnarray}
 H^2 &=& \frac{A_1}{(3-\beta^2)}a^{-3}\left[1+F\right]
 \\
  \rho_{\phi} &=& \frac{A_1}{(3-\beta^2)}a^{-3}\left[\beta^2+3F\right]
  \\
  p_{\phi} &=& -A_1a^{-3}F
  \\
  \omega_{\phi} &=& -(3-\beta^2)\left[\frac{F}{\beta^2+3F}\right] \\
  \Omega_{\phi} &=& \frac{1}{3}\left[\frac{\beta^2+3F}{1+F}\right] \\
  \Omega &=& \frac{1}{3}\left[\frac{3-\beta^2}{1+F}\right]
  \\
  q &=& \frac{1}{2}\left[\frac{1-(2-\beta^2)F}{1+F}\right]
  \end{eqnarray}
Apparently, fixing the present value of any one of the physical quantities will fix the same for other quantities for a given $\beta $. Hence, for example, fixing the value of $a_{0}/a_{t}$ from observational data one can estimate the value of $\beta$ such that it satisfactorily produces the present day value of other quantities. We can also plot (Fig.\ref{q_0vsbM1})the variation of those quantities with respect to $\beta$.

\par \noindent Plots show that present day values of the parameters are not much sensitive on $\beta$. Keeping in mind the uncertainty in observational data, the range of permissible vales of $\beta $ is thus not quite narrow. It may be interesting to note at this point that the model reduces to a $\Lambda$CDM one for $\beta =0$. Success of $\Lambda$CDM model predicts that $\beta$ will be close to zero.\\
The above model may be counted as one of many successful models that uses scalar fields as dark energy to explain late time acceleration. Let us now investigate whether this model can produce correct value of $a_0/a_{eq}$ by introducing radiation in it.

\section{Model 1A}
\begin{figure}
    \begin{center}
          \includegraphics[width=6cm]{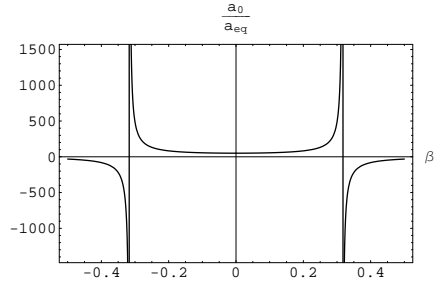}
  \caption{Variation of $a_0/a_{eq}$ with respect to $\beta$ with $a_0/a_t=1.7$ and $\Omega_{0}=0.28$ in Model 1A}\label{AeqvsbM1A}
  \end{center}
\end{figure}

Introduction of radiation in the Model 1 will make the background pressure $p\neq 0$ and hence we will need to make one assumption to solve the field equations. For reasons explained in section \ref{2} we choose the equation state to be
\begin{equation}
    \omega=\frac{1}{3}\left(1+\frac{a}{a_{eq}}\right)^{-1}.
\end{equation}
The conservation equation (\ref{conservation}) then leads to
\begin{equation}
    \rho=A_2a^{-4}\left(1+\frac{a}{a_{eq}}\right)
\end{equation}
and the field equations can now be solved again to give
\begin{equation}
    V(a)= \frac{A_2a^{-4}}{2(3-\beta^2)}\left[\left(\frac{2}{3}m+\frac{a}{a_{eq}}\right)\beta^2+(6-\beta^2)F_1\right]
\end{equation}
where $m$ and $F_1$ are defined as
\begin{eqnarray}
  m &=& \frac{3-\beta^2}{4-\beta^2} \nonumber\\
  F_1 &=& \left[\frac{8m}{3}\frac{a}{a_t}+\frac{a}{a_{eq}}\right]F.\nonumber
\end{eqnarray}
Other important physical quantities can now be listed as follows
\begin{eqnarray}
  H^2 &=& \frac{A_2a^{-4}}{(3-\beta^2)}\left[\left(\frac{4}{3}m+\frac{a}{a_{eq}}\right)+F_1\right] \\
  \rho_{\phi} &=& \frac{A_2a^{-4}}{(3-\beta^2)}\left[\left(m+\frac{a}{a_{eq}}\right)\beta^2+3F_1\right]  \\
  p_{\phi} &=& A_2a^{-4}\left[\frac{\beta^2}{3(4-\beta^2)}-F_1\right]  \\
  \omega_{\phi} &=& \frac{\frac{m}{3}\beta^2-(3-\beta^2)F_1}{\left(m+\frac{a}{a_{eq}}\right)\beta^2+3F_1} \\
  \Omega_{\phi} &=& \frac{1}{3}\left[\frac{\left(m+\frac{a}{a_{eq}}\right)\beta^2+3F_1}{\left(\frac{4}{3}m+\frac{a}{a_{eq}}\right)+F_1}\right] \\
  \Omega &=& \frac{1}{3}\left[\frac{\left(1+\frac{a}{a_{eq}}\right)(3-\beta^2)}{\left(\frac{4}{3}m+\frac{a}{a_{eq}}\right)+F_1}\right] \\
  q &=& \frac{\left(\frac{8m}{3}+\frac{a}{a_{eq}}\right)-(2-\beta^2)F_1}{\left(\frac{8m}{3}+\frac{2a}{a_{eq}}\right)+2F_1} \end{eqnarray}
One can see that Model 1A reduces to Model 1 in the limit $a\gg a_{eq}$ with the identification $A_1=A_2/a_{eq}$. It is equivalent of saying that late time evolution of the models are almost identical. But the added advantage in the Model 1A is that it is now capable of estimating the value of $z_{eq}$.
To see the variation of physical quantities with respect to $\beta$ we now have to fix the present day value of any two parameters, say $a_0/a_t$ and $\Omega_{0}$. Variations of $q_0$, $\omega_{\phi 0}$ with respect to beta are almost indistinguishable from those in Model 1 which is expected as late time evolution of both models are same and hence are not shown here. However, the most interesting thing to point out is that the value of $a_0/a_{eq}$ is highly sensitive with variation of $\beta$ in the permissible range(Fig.\ref{AeqvsbM1A}).

\begin{center}
\begin{tabular}{ccccc}
\hline
$\bf{\beta}$ &  &\bf{$z_{eq}$} & & \bf{$q_0$} \\\hline \hline
0.30000 & & 482.928 & & -0.534711 \\
0.31000 & & 1203.01 & &	-0.531834 \\
0.31200 & & 1728.83 & &	-0.531247 \\
0.31300 & & 2215.34 & &	-0.530952 \\
0.31400 & & 3086.80 & &	-0.530657 \\
0.31405 & & 3148.84 & &	-0.530642 \\
0.31410 & & 3213.43 & &	-0.530627 \\
0.31420 & & 3350.95 & &	-0.530597 \\
0.31430 & & 3500.81 & &	-0.530568 \\
0.31450 & & 3844.87 & &	-0.530508 \\
0.31500 & & 5099.15 & &	-0.530360 \\
\hline
\end{tabular}
\end{center}
\par

\noindent  Numbers exhibited in the table are calculated with $a_0/a_t=1.7$ and $\Omega_{0}=0.28$. It shows that a change in the third decimal place in the value of $\beta$ can change the value of $a_0/a_{eq}$ by few hundreds which means $\beta$ is now fixed more precisely in this model in view of accepted value of $z_{eq}=3196 \pm 134$ \cite{zeq}. This additional feature in Model 1A makes it much more capable than Model 1 to judge its viability.
\par Now the question is whether this sensitiveness of $a_0/a_{eq}$ on $\beta$ is a feature unique to this model or is it generic to all scalar field models?
In the next section, we investigate another model to seek answer to this question.

\section{Model 2}
Similar to Model 1 in this model also the matter is essentially dust and non-interacting hence the matter density again obeys the same equation
\begin{equation}
   \rho=A_3a^{-3}
\end{equation}
but we begin with a different assumption
\begin{equation}
    \dot{\phi}=\alpha a^n.
\end{equation}
Solution of the wave equation now becomes easy and the potential as a function of $a$ after dropping the constant part in it is given by
\begin{equation}
    V(a)=-\frac{\alpha^2}{2}\left(\frac{n+3}{n}\right)a^{2n}
\end{equation}
Using the condition that at $a=a_t$ the deceleration parameter goes to zero, a relation between $\alpha$ and $n$ can be found as
\[
\alpha^2=-\frac{A_3}{3}\left(\frac{n}{n+3}\right)a_{t}^{-(2n+3)}.
\]
Using this relation the potential can now be written as
\begin{equation}
    V(a)=\frac{A_3}{6}(n+3)G a^{-3}
\end{equation}
where the function $G$ is defined as
 \[
 G=(n+1)^{-1}\left(\frac{a}{a_t}\right)^{2n+3}.
 \]
Other physical quantities and parameters can now be listed as below
\begin{eqnarray}
  H^2 &=& \frac{A_3}{6}(2+G)a^{-3} \\
  \rho_{\phi} &=& \frac{A_3}{2}Ga^{-3} \\
  p_{\phi} &=& -\frac{A_3}{6}(2n+3)Ga^{-3} \\
  \omega_{\phi} &=& -\frac{1}{3}(2n+3) \\
  \Omega_{\phi} &=& \frac{G}{2+G} \\
  \Omega &=& \frac{2}{2+G} \\
  q &=& \frac{1-(n+1)G}{2+G}
\end{eqnarray}
The present value of the observable parameters can now be plotted against $n$ by fixing the value of $a_0/a_t$.
\begin{figure}
    \begin{center}
     \includegraphics[width=6cm]{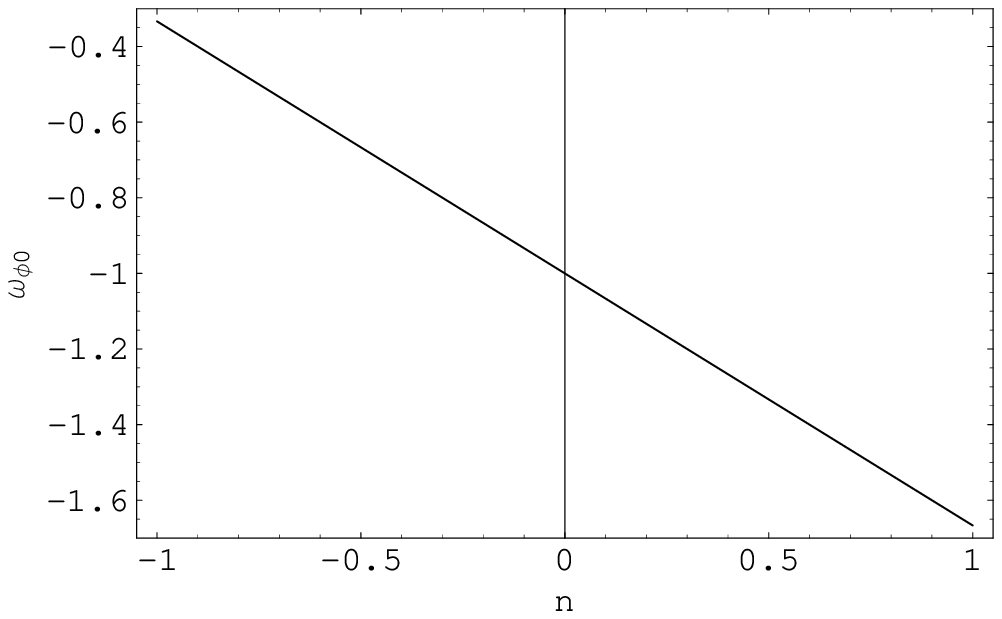}
     \includegraphics[width=6cm]{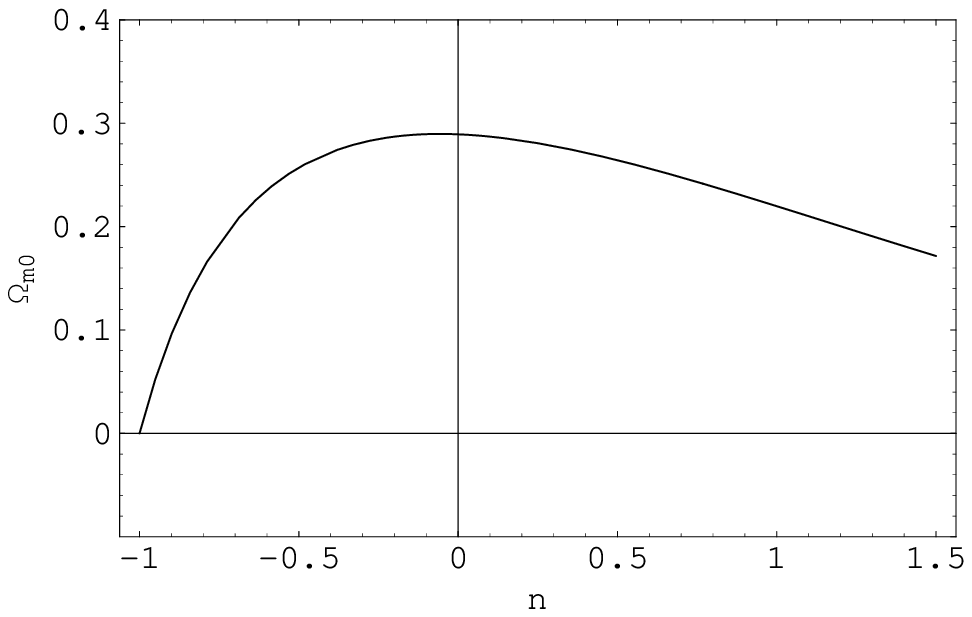}\\
     \includegraphics[width=6cm]{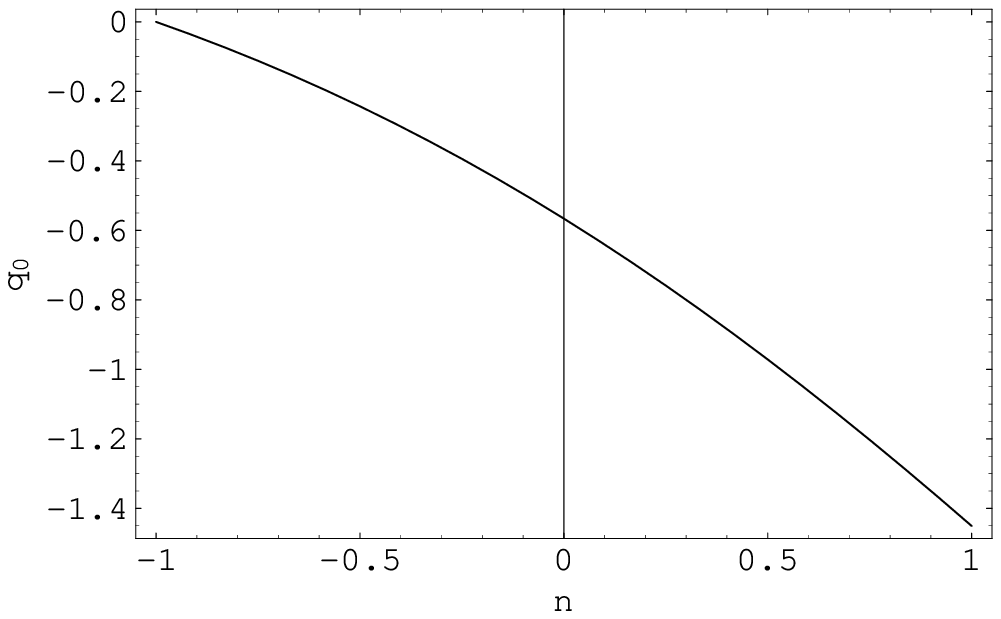}
     \includegraphics[width=6cm]{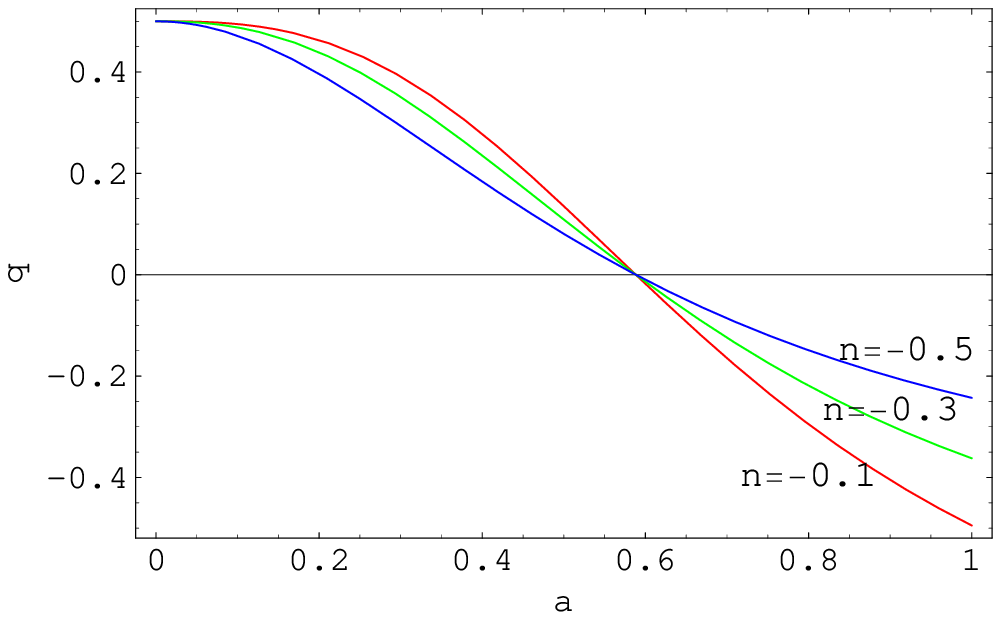}
  \caption{Variation of $\omega_{\phi0}$, $\Omega_{0}$ and $q_0$ with $n$ are shown for $a_{0}/a_{t} =1.7$ in Model 2 along with the nature of $q$ vs. $a$ curves for different values of $n $}\label{q_0vsnM2}
  \end{center}
\end{figure}
One can see a small negative value of $n$ is desirable to match the model with observation. A negative $n$ also fulfils the slow roll-over condition for $\phi$.
\par

To see whether this model can produce desirable value of $a_0/a_{eq}$ we introduce radiation in our next model following our prescription.

\section{Model 2A}
\begin{figure}
    \begin{center}
          \includegraphics[width=6cm]{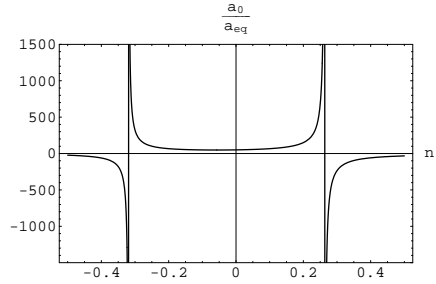}
  \caption{Variation of $a_0/a_{eq}$ with respect to $n$ with $a_0/a_t=1.7$ and $\Omega_{0}=0.28$ in Model 2A}\label{AeqvsnM2A}
  \end{center}
\end{figure}

Introducing radiation in the Model 2 will demand another assumption to solve the field equations. Similar to Model 1A in this model also we will assume
 \[
  \omega=\frac{1}{3}\left(1+\frac{a}{a_{eq}}\right)^{-1}
  \]
which leads to
\begin{equation}
    \rho=A_4a^{-4}\left(1+\frac{a}{a_{eq}}\right).
\end{equation}
The potential can now be obtained solving the wave equation. Using the condition that $q$ goes to zero at $a=a_t$ we have
\begin{equation}
   V(a)=\frac{A_4}{3}(n+3)G_1a^{-4}
\end{equation}
where $G_1$ is defined as
\[
G_1\equiv(n+1)^{-1}\left(1+\frac{1}{2}\frac{a_t}{a_{eq}}\right)\left(\frac{a}{a_t}\right)^{2n+4}.
\]
Other physical quantities and parameters solved from field equations are listed below
\begin{eqnarray}
  H^2 &=& \frac{A_4}{3}a^{-4}\left[\left(1+\frac{a}{a_{eq}}\right)+G_1\right] \\
  \rho_{\phi} &=& A_4G_1a^{-4} \\
  p_{\phi} &=& -\frac{A_4}{3}(2n+3)G_1a^{-4} \\
  \omega_{\phi} &=& -\frac{1}{3}(2n+3) \\
  \Omega_{\phi} &=& \frac{G_1}{\left(1+\frac{a}{a_{eq}}\right)+G_1} \\
  \Omega &=& \frac{\left(1+\frac{a}{a_{eq}}\right)}{\left(1+\frac{a}{a_{eq}}\right)+G_1} \\
  q &=& \frac{\left(1+\frac{1}{2}\frac{a}{a_{eq}}\right)-(n+1)G_1}{\left(1+\frac{a}{a_{eq}}\right)+G_1}
  \end{eqnarray}
As before the above equations reduce to their counterparts in Model 2 in the limit $a\gg a_{eq}$ and with the identification $A_3=A_4/a_{eq}$. Once again if we plot $a_0/a_{eq}$ vs. $n$ (Fig.\ref{AeqvsnM2A}), fixing two other observables like $a_0/a_t$ and $\Omega_{0}$ we find that the value of $a_0/a_{eq}$ is very sensitive with respect to changes in the value of $n$ and hence with changes in the present day values of other physical quantities.
\begin{center}
\begin{tabular}{ccccc}
\hline
\bf{${n}$} &  &\bf{$z_{eq}$} & & \bf{$q_0$} \\\hline \hline
-0.30000 &  & 319.988 &  &	-0.363535 \\
-0.31000 &  &	654.358 &  &	-0.356587 \\
-0.31400 &  &	1150.13 &  &	-0.353798 \\
-0.31500 &  &	1422.49 &  &	-0.353102 \\
-0.31720 &  &	2991.89 &  &	-0.351569 \\
-0.31725 &  &	3069.22 &  &	-0.351534 \\
-0.31730 &  &	3150.67 &  &	-0.351500 \\
-0.31740 &  &	3327.32 &  &	-0.351430 \\
-0.31745 &  &	3423.32 &  &	-0.351395 \\
-0.31750 &  &	3525.05 &  &	-0.351360 \\
-0.31800 &  &	5017.90 &  &	-0.351012 \\
\hline
\end{tabular}
\end{center}

The numbers displayed in the table are once again calculated with $a_0/a_t=1.7$ and $\Omega_{0}=0.28$.
Thus it appears that sensitiveness of $a_0/a_{eq}$ on present day values of observable quantities may be generic to these kind of scalar field models.
\section{Conclusion}
The present article is an attempt to make dark energy models capable of estimating $z_{eq}$ by introducing an effective equation of state for radiation-matter. Two cosmological models (Model 1 and Model 2) with a minimally coupled scalar field playing the role of dark energy are chosen to demonstrate this possibility. These models are no different from other successful models existing in literature. One may find it difficult to judge the better one in the light of present day observation. Including radiation in these models by introducing a new equation of state we arrive at Model 1A and Model 2A respectively. The late time behaviors of these new models remain unchanged. To judge which one among these new models is better we can now estimate $z_{eq}$ from those and compare with accepted values in literature. To our surprise we find that estimated values of $z_{eq}$ from these models are too sensitive with respective values of parameters present in these models. In other words the values of observable quantities are now fixed up to third decimal of places in order to produce correct $z_{eq} $.
Thus inclusion of the proposed new equation of state makes dark energy models capable of estimating $z_{eq}$ and can be used as a new method for judging the merit of a dark energy model.

\section{Acknowledgement}
One of the authors, A. Sil wishes to thank University Grants Commission, Eastern Regional Office at Kolkata for financial assistance by approving a Minor Research Project (F. PSW-050/08-09 ERO SNO.88638)to carry out the present work.

\end{document}